\documentclass[article,twocolumn,floatfix]{revtex4}

\usepackage{graphicx} 
\usepackage{subfig}
\usepackage{color}

\newcommand{\ket}[1]{\left| #1 \right\rangle}
\newcommand{\bra}[1]{\left\langle #1 \right|}

\newcommand{\be}{\begin{equation}}
\newcommand{\ee}{\end{equation}}
\newcommand{\bea}{\begin{eqnarray}}
\newcommand{\eea}{\end{eqnarray}}

\usepackage{dcolumn}
\usepackage{bm}
\usepackage{amssymb,amsmath}
\usepackage{color}
\usepackage{float}
\usepackage{tikz}
\usetikzlibrary{arrows}

\definecolor{DarkGreen}{rgb}{0,0.6,0.2}

\begin{document}
\title{
Single-photon Spectra  in Quantum Optomechanics}
\author{Imran M. Mirza, S.J. van Enk}
\affiliation{Oregon Center for Optics and Department of Physics\\University of Oregon\\
Eugene, OR 97403}
\begin{abstract}
We consider how a single photon can probe the quantum nature of a moving mirror in the context of quantum optomechanics. In particular, we demonstrate how the single-photon spectrum reveals resonances that depend on how many phonons are created as well as on the strength of the mirror-photon interaction. A dressed-state picture 
is used to explain positions and relative strengths of those resonances. The time-dependent spectrum shows how the resonances are built up over time by the photon interacting with the moving mirror.
\end{abstract}
  \maketitle
\section{Introduction}
Quantum optomechanics has become an active area of research in which the quantized center-of-mass motion of a tiny mirror plays the central role. 
One aim is to probe the existence of coherent superpositions of macroscopically different quantum states and study their decay mechanisms \cite{marshall2003towards,kleckner2008creating}. A quantum mirror may also be used for quantum information processing purposes, e.g., to
store optical information in the mechanical motion
of the mirror. By mapping this information back into a different optical mode, frequency conversion
of photonic quantum states may be achieved \cite{tian2010optical}.

So far, the quantized motion has been studied experimentally by means of laser light interacting with the moving mirror. 
Here we study the possibilities that arise from having a non-negligible interaction between a single photon and a mirror (a mildly futuristic possibility, but one that starts being taken very seriously \cite{nunnenkamp2011single,pepper2012optomechanical,hammerer2012nonclassical,akram2013,sekatski2014macroscopic}). 

In particular, we will describe theoretically how the {\em time-dependent  spectrum} of a single photon is modified by its interaction with a moving mirror. The time-dependent spectrum of a single photon is defined as follows: 
Imagine that a single-photon wavepacket enters a (Lorentzian) filter cavity, described by a resonance frequency $\omega$ and a filter bandwidth $\Gamma$. We can record as a function of time when the photon exits the filter. If we represent the detuning between filter resonance and optomechanical cavity frequency $\omega_{c}$ by $\Delta=\omega-\omega_{c}$, then we can express the expected counting rate at time $t$ in terms of the continuous field annihilation and creation operators $\hat{a}(t)$ and  $\hat{a}^\dagger(t)$ of the photon wavepacket as
\begin{equation}\label{TDPS}
\begin{split}
&N(t;\Delta,\Gamma)=\Gamma^{2}\int_{0} ^t\int_{0} ^t e^{-(\Gamma-i\Delta)(t-t^{'})}e^{-(\Gamma+i\Delta)(t-t^{''})}\times\\
&\hspace{20mm}\langle\hat{a}^{\dagger}(t^{'})\hat{a}(t^{''})\rangle dt^{'}dt^{''}.
\end{split}
\end{equation}
This is the time-dependent spectrum (it depends both on $\omega$ and $t$). It has the same form as that introduced by Eberly and Wodkiewicz \cite{eberly1977time}. The main difference is that the classical field amplitudes in the Eberly and Wodkiewicz spectrum are here replaced by quantum annihilation and creation operators.

We can also integrate the time-dependent spectrum over time, and thus define  
 \begin{equation}\label{SPS}
 N_S(t;\Delta,\Gamma)=
 \int_0^t N(t';\Delta,\Gamma)dt'.
 \end{equation}
 In the limit of $t\rightarrow\infty$ this quantity would equal the spectrum for a stationary process as obtained from the Wiener-Khinchine theorem \cite{mandel1995optical}, which is perhaps the more familiar quantity.

We are going to use two theoretical methods in the following. First, the whole process of detecting
a single photon emanating from a cavity is very well described by the quantum trajectory method \cite{molmer1993monte, dum1992monte, wiseman1996quantum, plenio1998quantum,carmichael2008statistical}, especially when combined with input-output theory \cite{gardiner2004quantum, ciuti2006input}.
As we have shown before \cite{mirza2013single}, the time-dependent single photon spectrum (as well as its infinite-time limit) can be straightforwardly calculated using the method developed in \cite{carmichael2008statistical}. Second, we find that a simple dressed-state picture 
suffices to understand the locations and heights of the resonances that become visible in these spectra.

\section{Theoretical description}
\begin{figure*}[t]
\hspace{-20mm}\includegraphics[width=4.4in]{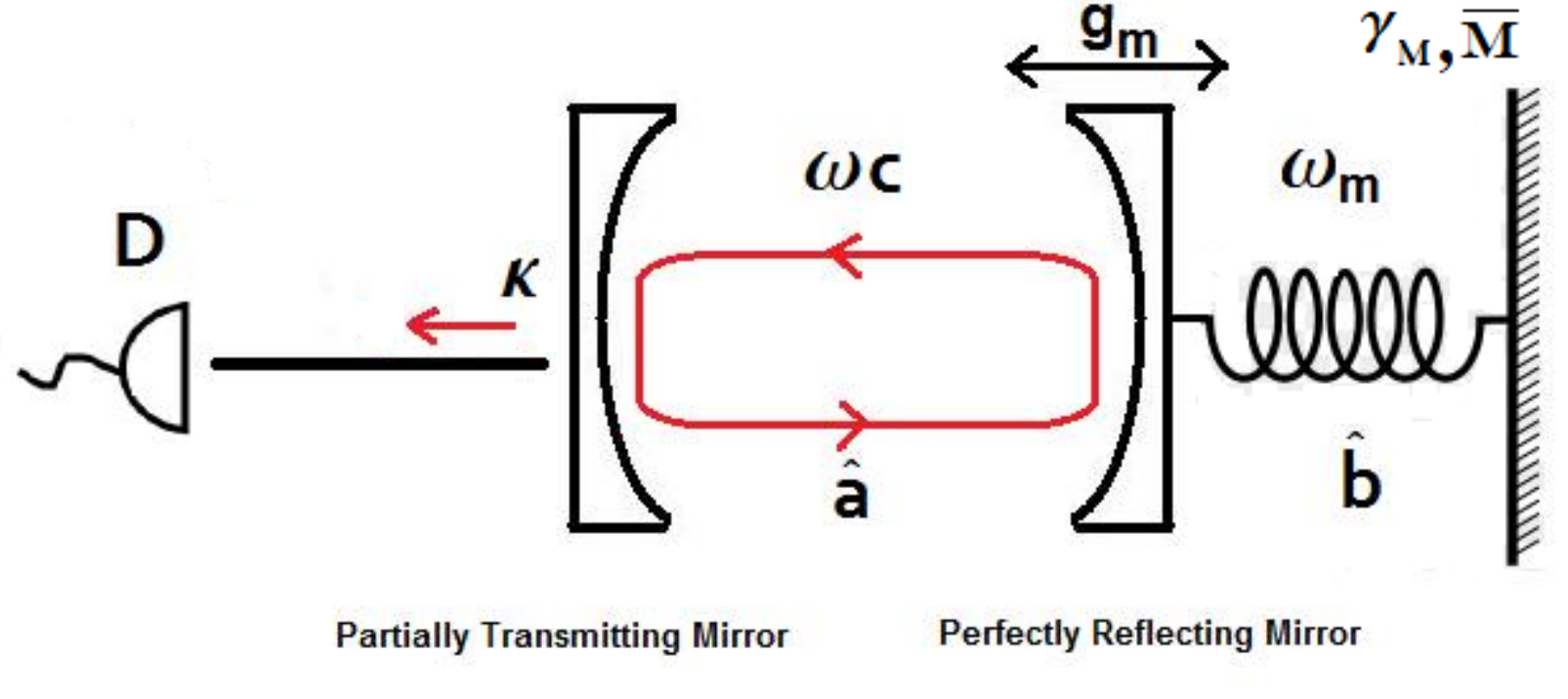}
\captionsetup{
  format=plain,
  margin=1em,
  justification=raggedright,
  singlelinecheck=false
}
  \caption{An optomechanical cavity (OMC): A Fabry-Per\'ot cavity with a fixed mirror on the left and a movable mirror on the right. Initially the OMC is assumed to contain a single photon and zero phonons. The right mirror (modeled as a harmonic oscillator) is coupled to a finite temperature heat bath with mechanical decay rate $\gamma_{M}$, while the temperature of the bath is given in terms of the average thermal phonon number $\overline{M}$. The decay rate of the optical cavity field amplitude (due to leakage of photons through the left mirror) is denoted by $\kappa$. The output field is detected and its time-dependent spectrum is used to probe the motion of the movable mirror.}\label{Fig1}
\end{figure*}
As shown in FIG.~1 we consider an optomechanical cavity (OMC) system \cite{aspelmeyer2013cavity, kippenberg2008cavity, park2009resolved}. The OMC is a Fabry-P\'erot cavity with a perfectly reflecting movable right mirror, which is modeled as a quantum harmonic oscillator with frequency of oscillation $\omega_{M}$. The annihilation of quanta of mechanical vibrations (phonons) is described by the operator $\hat{b}$. Initially the OMC is assumed to have a single photon inside, and we assume  that only a single resonant optical mode is relevant. The frequency of that mode is $\omega_{c}$,  and annihilation of photons in that mode is described by the operator $\hat{a}$. We study the situation where the right mirror is so thin that even a single photon can affect its mechanical motion through radiation pressure \cite{jackson1998classical, girvin2009trend}. The rate describing the coupling strength between optical and mechanical degrees of freedom is denoted by $g_{M}$. 

We shall start by writing down the standard Hamiltonian \cite{aspelmeyer2013cavity, milburn2011introduction, meystre2012short} of such an  OMC while leaving out the zero-point energies of the oscillators, so that
\begin{equation}\label{Hsys}
\hat{H}_{sys.} = \hbar\omega_{c}\hat{a}^{\dagger}\hat{a}+\hbar\omega_{M}\hat{b}^{\dagger}\hat{b}-\hbar g_{M}\hat{a}^{\dagger}\hat{a}(\hat{b}+\hat{b}^{\dagger}).
\end{equation}
The  left mirror is taken to be partially transmitting and hence  the photon wavepacket can leak out through that left mirror. We assume this happens at a rate $\kappa$ and that the photon wave packet enters an optical fiber (which is assumed to have a continuum of modes). The fiber is introduced just for the purpose of ensuring unidirectional propagation of the emitted photon wavepacket towards a detector. The transmission, reflection and emission spectra of a single photon wavepacket from either a one-sided or a two-sided Fabry-P\'erot cavity (with both mirrors fixed) have all been very well studied \cite{ujihara2010output, saleh1991fundamentals}, but here we address the  situation where one of the mirrors of the OMC is capable of harmonic motion.

Due to the presence of the continuum of modes in the fiber, our problem is essentially an open quantum system problem. There are several different approaches which can be applied in order to calculate the spectrum emitted by such a system \cite{carmichael1993open, breuer2002theory, weiss2012quantum}. Here we will make use of the Quantum Jump/Trajectory (QJT) approach \cite{molmer1993monte, dum1992monte, wiseman1996quantum, plenio1998quantum,carmichael2008statistical} combined with input-output theory \cite{gardiner2004quantum, ciuti2006input}. 

There is one output detector in our system and we introduce the corresponding output operator $\hat{J}_{out}$ (which is basically the continuous-mode annihilation operator for the output field). This output operator  is related to the input operator $\hat{J}_{in}$  through the standard input-output relationship \cite{gardiner2004quantum} as: 
\begin{equation}\label{JO}
\hat{J}_{{\rm out}}(t)=\hat{J}_{{\rm in}}(t)+\sqrt{\kappa}\hat{a}(t)
\end{equation}
where we have neglected the trivial fiber time delays between the OMC and the detector (following the standard cascaded quantum jump approach \cite{gardiner1993driving}). We also can often disregard the input operator in the above equation as it is not going to contribute to normally ordered observables (all our observables of interest are of that form). This is because if we denote by $\ket{\Psi}$ the initial state of the global system (cavity and fiber) we have $\hat{J}_{in}\ket{\Psi} = 0$ as we assume there is no (fiber) photon present to serve as input for the OMC. 
\section{Single-phonon mechanical oscillations}
In this section, we consider a rather simple situation where we assume that initially there is no phonon present and the photon inside the OMC will generate at most a single phonon.
\subsection{Absence of mechanical losses and zero temperature}
We first consider the simplest case of no mechanical losses and zero temperature for the mechanical heat bath. That is,   we assume a zero phonon leakage rate ($\gamma_{M} = 0$), and we assume the average thermal phonon number $\overline{M}$ to be zero. Note that the first condition can always be physically realized for short times $t$ such that 
\[
\frac{1}{\kappa}<t<\frac{1}{\gamma_{M}}.
\] 
In the next subsection we will discuss the effects of mechanical losses and nonzero temperature on the behavior of the spectrum.

According to the QJT approach we have the following picture. In any given small time interval we have one of two
situations:

(I) Occurrence of a quantum jump: whenever our detector records a click, a quantum jump takes place, and we apply the output (annihilation) operator $\hat{J}_{out}$ (which in this context is also termed the ``jump operator'')
to the state of the system.

(II) No quantum jump takes place: while no detector clicks, the system evolves according to the following non-Unitary Schr\"odinger equation:
\begin{equation}\label{NUSE}
i\hbar\frac{d}{dt}\ket{\tilde{\psi}(t)}=\hat{H}_{NH}\ket{\tilde{\psi}(t)},
\end{equation}
where the ``Hamiltonian'' appearing in this equation is a non-Hermitian operator, which is the sum of two parts. The first part is Hermitian and is given by the system's Hamiltonian, the second part is anti-Hermitian and  is constructed from the jump operator. In total we have
\begin{equation}\label{NHH}
\begin{split}
&\hat{H}_{NH}=\hat{H}_{sys}-i\hbar\hat{J}^{\dagger}_{out}\hat{J}_{out}\\
&=\hbar\omega_{c}\hat{a}^{\dagger}\hat{a}+\hbar\omega_{M}\hat{b}^{\dagger}\hat{b}-\hbar g_{M}\hat{a}^{\dagger}\hat{a}(\hat{b}+\hat{b}^{\dagger})-i\hbar\frac{\kappa}{2}\hat{a}^{\dagger}\hat{a}.
\end{split}
\end{equation}
Restricting the phonon number to be at most 1 (one) we can also define a mechanical annihilation operator as $\hat{b}=\ket{0}_{b}\bra{1}$, in terms of the states $\ket{0}_{b}$ and $\ket{1}_{b}$, the zero- and one-phonon number states, respectively. This restriction also implies that $\hat{b}^{\dagger}\ket{m}=0$, $\forall$ $m\geq 1$. Notice that with this new notation the Hermitian nature of $\hat{H}_{sys}$ will not be disturbed. 

The unnormalized state $\ket{\tilde{\psi}(t)}$ appearing in Eq. [\ref{NUSE}] is called the ``No-jump state'' in QJT and it is a superposition of all the different possibilities of finding the excitation in the system before it is being lost by the system and registered by the detector. For our setup it can be written as: 
\begin{equation}\label{NJS}
\ket{\tilde{\psi}(t)} = c_{1}(t)\ket{10}+c_{2}(t)\ket{11},
\end{equation}
where we use the following  notational convention: the first place in the ket gives the number of photons in the OMC and the second place the number of phonons in the mechanical oscillator. The probability amplitudes appearing in the No-jump state can easily be worked out by using Eqs.[\ref{NJS}] and [\ref{NHH}] in Eq.~[\ref{NUSE}]. In Laplace space these amplitudes then turn out to be
\begin{subequations}
\begin{eqnarray}
C_{1}(s) = \Bigg[\frac{s+\frac{\kappa}{2}+i\omega_{M}}{(s+\frac{\kappa}{2})(s+\frac{\kappa}{2}+i\omega_{M})+g^{2}_{M}}\Bigg],\\
\label{C1s}
C_{2}(s) = \Bigg[\frac{ig_{M}}{(s+\frac{\kappa}{2})(s+\frac{\kappa}{2}+i\omega_{M})+g^{2}_{M}}\Bigg],\label{C2s}
\end{eqnarray}
\end{subequations}
where $C_{i}(s)$ is the Laplace transform of $c_{i}(t)$ with $i = 1, 2$. 

For the spectrum calculations in QJT,  we first have to calculate the time-dependent spectrum and then by taking the $t\rightarrow\infty$ limit we can obtain the infinitely long time spectrum. We use the equations given in the Introduction, with the generic operator $\hat{a}$ replaced by $\hat{J}_{out}$. The latter is determined by Eq.~[\ref{JO}], and
expectation values involving $\hat{a}$ and $\hat{a}^\dagger$ are determined by the coefficients $C_i(s)$.
Our infinitely long time spectrum $P_{D}({\Delta,\infty})$, can be evaluated analytically, and the result is
\begin{equation}\label{ILTS}
\begin{aligned}
&\hspace{-5mm}P(\Delta,\infty)=\kappa\Gamma\Bigg(\Bigg|C_{1}(s=-i\Delta)\Bigg|^{2}+\Bigg|C_{2}(s=-i\Delta-i\omega_{M})\Bigg|^{2} \Bigg)\\
&\hspace{8mm}=\kappa\Gamma\Bigg[\Bigg|\frac{i(\omega_{M}-\Delta)+\kappa/2}{\lbrace i(\omega_{M}-\Delta)+\kappa/2 \rbrace \lbrace \kappa/2-i\Delta \rbrace+g^{2}_{M}}\Bigg|^{2}\\
&\hspace{10mm}+\Bigg|\frac{ig_{M}}{\lbrace \kappa/2-i(\omega_{M}+\Delta)\rbrace \lbrace \kappa/2-i\Delta \rbrace+g^{2}_{M}}\Bigg|^{2}\Bigg].
\end{aligned}
\end{equation}
This is the main result for this subsection. We plot this spectrum  in various different regimes:\\ 
\begin{figure*}
\hspace{7mm}\vspace{-5mm}\text{{\bf (a)}}\hspace{80mm}\vspace{-2mm}\text{{\bf (b)}}
\begin{center}
\begin{tabular}{cccc}
\subfloat{\includegraphics[width=3in,height=6in]{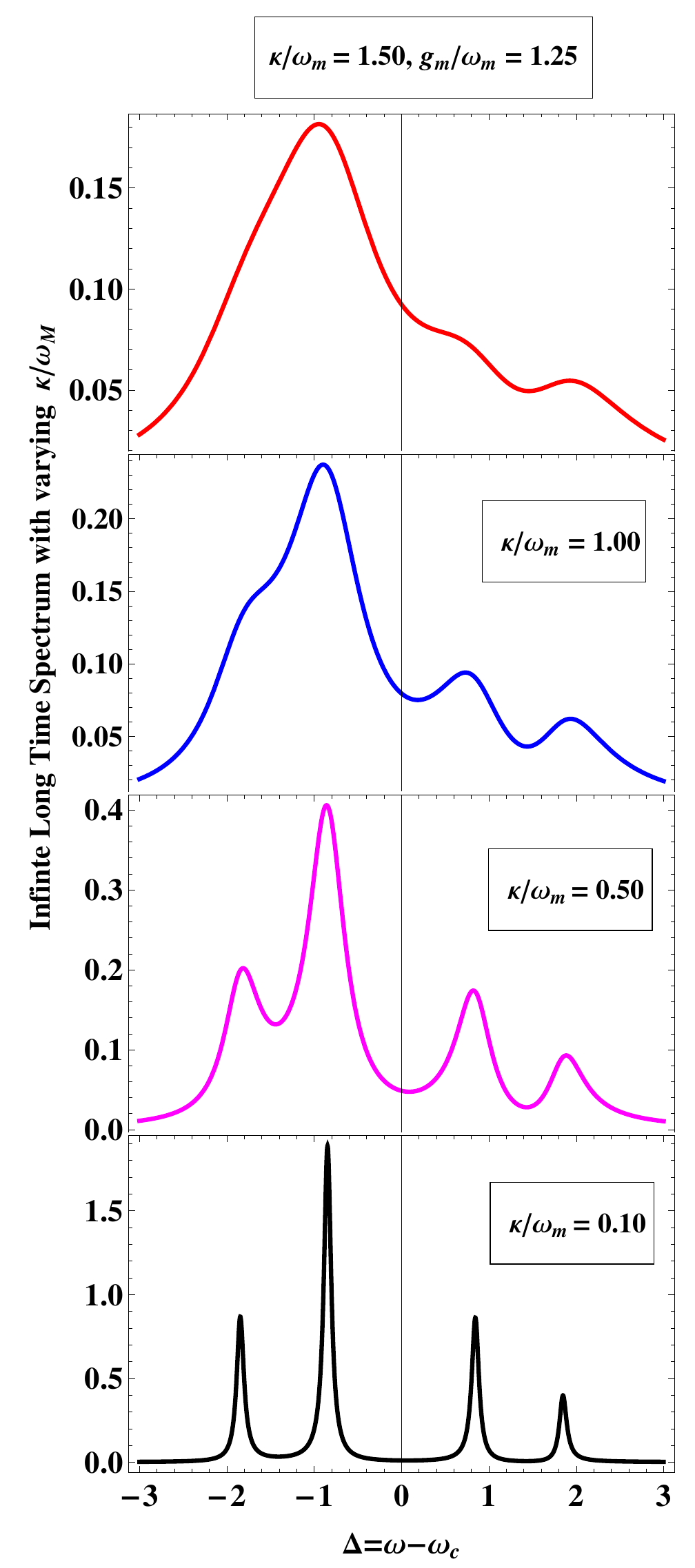}}&
\hspace{10mm}\subfloat{\includegraphics[width=3in,height=6in]{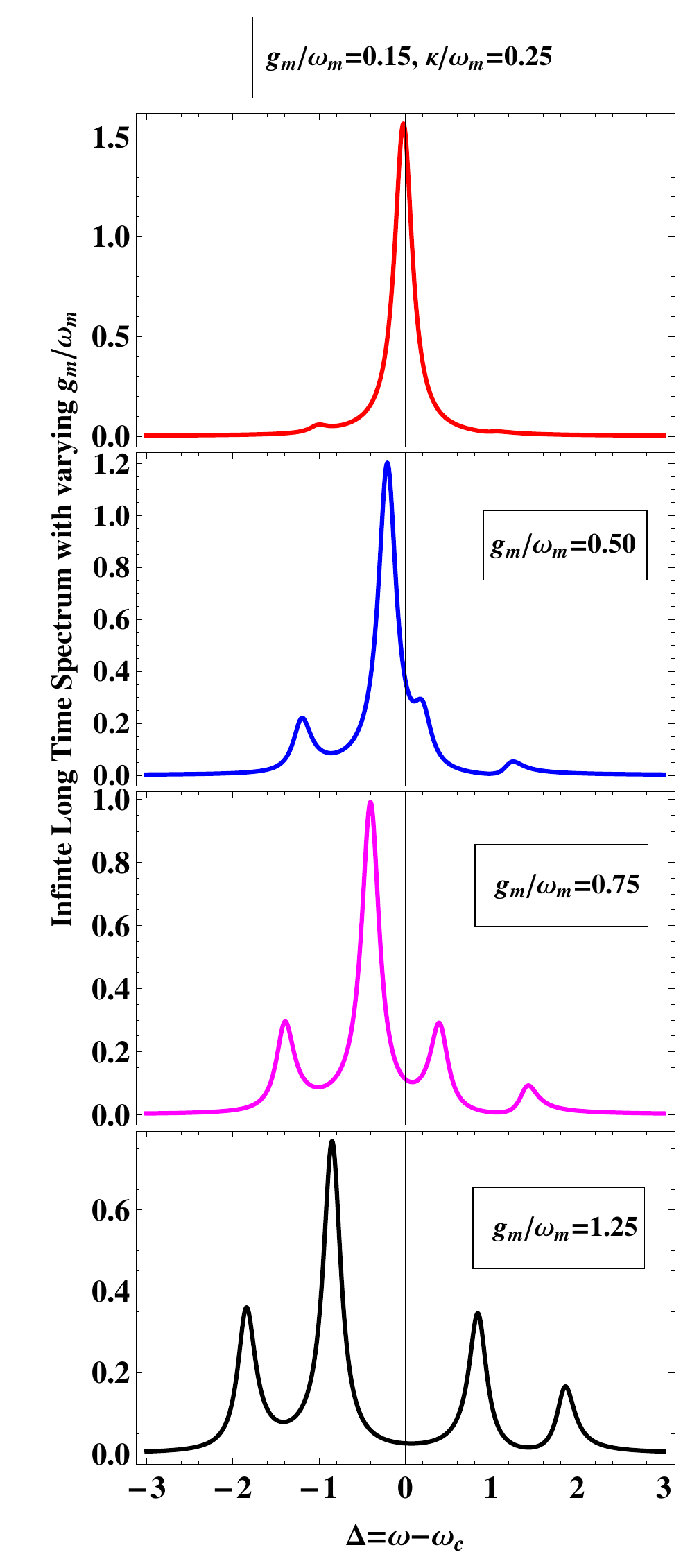}}\\
\end{tabular}
\captionsetup{
  format=plain,
  margin=1em,
  justification=raggedright,
  singlelinecheck=false
}
\caption{Single photon spectrum in the limit $\rightarrow\infty$ as emitted by the OMC, in the case that there is at most one phonon present. 
The common parameters used in both parts (a) and (b) are the frequency filter parameters $\Delta/\omega_{M} = 0.5$ and $\Gamma/\omega_{M}=0.1$.
{\bf (a)} Comparison of the good cavity ($\kappa<\omega_{M}$) and bad cavity ($\kappa > \omega_{M}$) regimes. From top to bottom curves one sees that side bands in the spectrum become better and better resolved. In all curves $g_{M}/\omega_{M} = 1.25$.
 {\bf (b)} Varying the values of $g_{M}$ as compared to $\kappa$ (in units of $\omega_{M}$) so that a comparison between the strong and weak coupling regimes can be made. For all curves we have $\kappa/\omega_{M} = 0.25$. 
In both parts (a) and (b) of the figure  the separation between the two middle peaks is determined by $g_{M}$ while the two side peaks are $\omega_{M}$ farther away from the middle peaks.}\label{Fig2}
\end{center}
\end{figure*}
{\bf Good and Bad cavity limits:} First we plot this spectrum in the good ($\kappa < \omega_{M}$) and bad ($\kappa > \omega_{M}$) cavity limits \cite{marquardt2007quantum, marquardt2009optomechanics}. FIG.~[2(a)] shows our results. For the parameters chosen (especially the value of $g_{M}/\omega_{M}$ ) we note that even in the bad cavity limit there are some resonant structures visible, even though all resonances are overlapping. Gradually going to the good cavity limit and finally approaching $\kappa = 0.10 \omega_{M}$ we see that all resonances are now separated and can easily be distinguished from one another.\\
{\bf Strong and weak coupling regimes:} Next we vary $\kappa$ and $g_{M}$ in units of $\omega_{M}$. In FIG.~[2(b)] we have plotted our results in both weak $(\kappa>g_{M})$ and strong $(\kappa<g_{M})$ coupling regimes \cite{groblacher2009observation, weiss2013strong}. 
We note that starting from the weak coupling regime (top red curve in the figure) we have just one major peak centred at the optical resonance frequency and a very tiny shoulder (side-band) on the left side of the peak.  This left side shoulder (the ``red side band'' \cite{chan2011laser}) indicates a process in which the photon first produces a phonon and then leaves the cavity with a frequency smaller than $\omega_c$.  
 For the red curve, the red side band is very small because the coupling between the single photon and the single phonon is kept small compared to the photon escape rate $\kappa$. In the next three curves (blue, pink and black) we enter into the strong coupling regime. With this change we start to observe all side bands clearly. Specifically, we now see the appearance of blue side bands as well (on the positive side of the $\Delta$ axis). These new side bands refer to the processes in which the photon first produces a phonon and then takes energy from away from that phonon and leaves the cavity with energy greater than $\hbar\omega_c$.  

From this figure we can conclude that a fully resolved OMC spectrum (the black curves in FIG.~2) can be observed if we work in the strong coupling regime within the good cavity limit.
\subsubsection{Locations of the resonances}
The locations of the resonances can be worked out by setting the real part of the poles in the spectrum of Eq.~[\ref{ILTS}]
equal to zero. There are two terms in the spectrum and each term gives us two resonances (poles).  We thus find four resonances determined by
\begin{equation}
\Delta=\pm\frac{\omega_{M}}{2}\pm\sqrt{\frac{\omega^{2}_{M}+\kappa^{2}}{4}+g^{2}_{M}}.
\end{equation}
The locations of the resonances can also be worked out by performing a dressed-states analysis \cite{barnett2002methods} of this problem. 

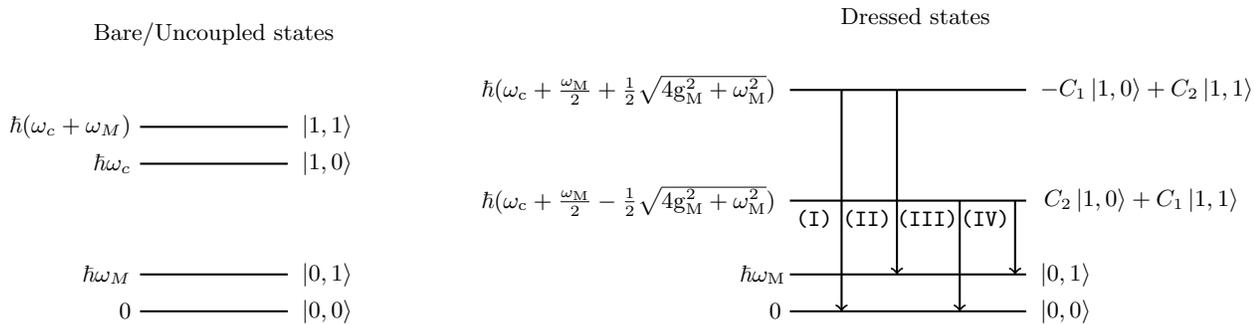
\begin{figure*}
\hspace*{-3cm}%
\centerline{
  \resizebox{15cm}{!}{
    \begin{tikzpicture}  
     \draw[thick](-5,0) -- (-3,0) node at (-5,0)[left] {{$0$}}node at (-2,0)[left] {{$\ket{0,0}$}};
     \draw[thick](-5,2) -- (-3,2) node at (-5,2)[left] {{$\hbar \omega_{c}$}}node at (-2,2)[left] {{$\ket{1,0}$}};
     \draw[thick](-5,2.5) -- (-3,2.5) node at (-5,2.5)[left] {{$\hbar (\omega_{c}+\omega_{M})$}}node at (-2,2.5)[left] {{$\ket{1,1}$}};
     \draw[thick](-5,0.5) -- (-3,0.5)node at (-5,0.5)[left] {{$\hbar\omega_{M}$}}node at (-2,0.5)[left] {{$\ket{0,1}$}}node at (-4,3.75){{{$ \small \text {Bare/Uncoupled states}$}}}node at (5.5,4){{{$ \small \text{Dressed  states}$}}};        
    \hspace{20mm}\draw[thick](1.8,0) -- (5,0) node at (6,0)[left]{$\small{\ket{0,0}}$}node at (1.4,0)[right]{$0$};
                 \draw[thick](1.8,0.5) -- (5,0.5)node at (6,0.5)[left]{$\small {\ket{0,1}}$}node at (1,0.5)[right]{${\rm \hbar\omega_{M}}$};            
                 \draw[thick](1.8,1.5) -- (5,1.5)node at (8,1.5)[left]{$\small {C_{2}\ket{1,0}+C_{1}\ket{1,1}}$}node at (-2.5,1.5)[right]{${\rm \hbar(\omega_{c}+\frac{\omega_{M}}{2}-\frac{1}{2}\sqrt{4g_{M}^{2}+\omega_{M}^{2}})}$};
                 \draw[thick](1.8,3) -- (5,3)node at (8.2,3)[left]{$\small {-C_{1}\ket{1,0}+C_{2}\ket{1,1}}$}node at (-2.5,3)[right]{${\rm \hbar(\omega_{c}+\frac{\omega_{M}}{2}+\frac{1}{2}\sqrt{4g_{M}^{2}+\omega_{M}^{2}})}$};
                  \draw[thick][->](2.5,3) -- (2.5,0)node at (2.5,1.25)[left]{$\texttt{(I)}$};
                  \draw[thick][->](3.25,3) -- (3.25,0.5)node at (3.3,1.25)[left]{$\texttt{(II)}$};
                  \draw[thick][->](4.1,1.5) -- (4.1,0)node at (4.2,1.25)[left]{$\texttt{(III)}$};
                  \draw[thick][->](4.85,1.5) -- (4.85,0.5)node at (4.9,1.25)[left]{$\texttt{(IV)}$};
    \end{tikzpicture}
    }
    }
\captionsetup{
format=plain,
margin=1em,
justification=raggedright,
singlelinecheck=false
}
\caption{A energy level diagram explaining the appearance of dressed states as a result of coupling between bare optomechanical states. Different transitions are numbered corresponding to spectrum peaks in FIG.~4}
\end{figure*}
Diagonalizing the system Hamiltonian (Eq.[\ref{Hsys}]) using $\lbrace \ket{100},\ket{110},\ket{001},\ket{011}\rbrace$ as a basis, produces the set of eigenvalues: 
\begin{equation}
\begin{aligned}
& \Bigg\lbrace 0,\hspace{3mm}\hbar\omega_{M},\hspace{3mm} \hbar\Bigg(\omega_{c}+\frac{\omega_{m}}{2}\pm\frac{1}{2}\sqrt{4g_{M}^{2}+\omega_{M}^{2}}\Bigg)\Bigg\rbrace.
\end{aligned}
\end{equation}
The corresponding eigenvectors $\ket{\lambda_{i}}$, $\forall$ $1\leq i\leq4$ (the so-called dressed states) take the form
\begin{subequations}
\begin{eqnarray}
\ket{\lambda_{1}}=\ket{0,0},\ket{\lambda_{2}}=\ket{0,1},\\
\ket{\lambda_{3}}=-N_{1}\ket{1,0}+N_{2}\ket{0,1},\\
\ket{\lambda_{4}}=N_{2}\ket{1,0}+N_{1}\ket{0,1},
\end{eqnarray}
\end{subequations} 
with  
\begin{equation}
N_{1}=\frac{2g_{M}}{\sqrt{4g_{M}^{2}+(\omega_{M}+\sqrt{4g_{M}^{2}+\omega_{M}^{2}})^{2}}},
\end{equation}
 and 
 \begin{equation}
 N_{2}=\frac{(\omega_{M}+\sqrt{4g_{M}^{2}+\omega_{M}^{2}})}{\sqrt{4g_{M}^{2}+(\omega_{M}+\sqrt{4g_{M}^{2}+\omega_{M}^{2}})^{2}}}.
 \end{equation} 
The energy level diagram showing the dressed states resulting from the optomechanical coupling between bare states is shown in FIG.~3. We note that because of the coupling between optical and mechanical degrees of freedom, the two bare states having a single photon combine to form two dressed states (upper two states in right part of the figure). In the dressed states picture there are  four transitions  possible among the different states. All of these four transitions have a different frequency. And these frequencies turn out to be exactly located at the peak locations in FIG.~2  worked out above by setting the real part of the poles zero.

There is another and rather simpler way of expressing the resonances. For that we can summarize the peak locations as $\Delta=\frac{g^{2}_{M}}{\omega_{M}}-m\omega_{M}$ for $m=0,1$ for both positive and negative axes of $\Delta$. This expression is just a compact form of writing the peak positions by looking at the eigenvalues of the system Hamiltonian which we stated above, and it is consistent with results reported in \cite{nunnenkamp2011single}.

\subsubsection{Asymmetry in the peak heights} We note that two of the peaks in the fully resolved spectrum are of equal height but the other two are asymmetric. Asymmetric and symmetric peaks are associated with the first and the second terms in the emission spectra of Eq.(\ref{ILTS}), respectively. We notice that mathematically this asymmetry can be attributed to the presence of the detuning parameter $\Delta$ in the numerator of first term, which under the exchange $\Delta\longleftrightarrow -\Delta$ breaks the symmetry in the heights. 

 We can further explain this asymmetry by looking at the dressed state picture of the problem. Once the photon interacts with the mechanical motion it can either create a single phonon or no phonons at all. Corresponding to these two choices, the system can be either in one dressed state $\ket{\lambda_{3}}$ or in the other one, $\ket{\lambda_{4}}$. In general  we can express the state of the system as a superposition of both these dressed states. This superposition is in general imbalanced and hence the photonic transitions from such a superposed state to the two lower states ($\ket{\lambda_{1}},\ket{\lambda_{2}}$) will lead to unequal peak heights.
 
This explanation predicts, in fact, that in general all peaks should be of different heights. We confirmed this prediction by varying the system parameters: we found that indeed the symmetry in the two peaks of the spectrum as shown in FIG.~2 (bottom plots) is accidental and is not always present (as Figures 5 and 6 below will confirm). Moreover, we expect that, by starting with one phonon present in the initial state, these peaks should interchange their heights, as in that case the transitions between the one-phonon excited state to the one-phonon ground state will become the main process. The plot in FIG.~4 confirms this expectation.
\begin{figure}
    \includegraphics[width=3in,height=2in]{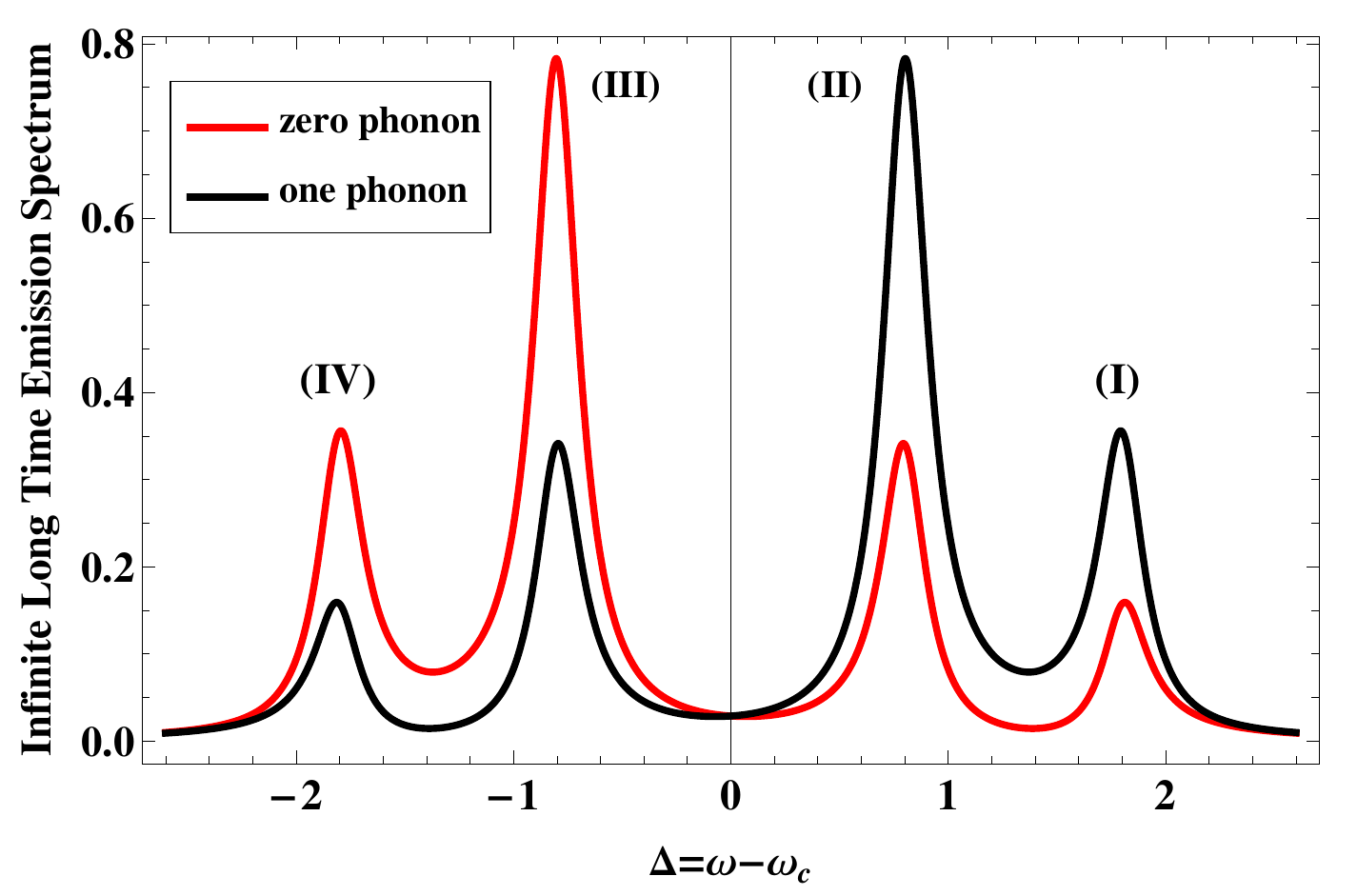}
    \captionsetup{
  format=plain,
  margin=1em,
  justification=raggedright,
  singlelinecheck=false
}
 \caption{Effect of changing the initial phonon number in the OMC from zero to one on the long time spectrum. The parameters used are: $\kappa/\omega_{M}=0.25$, $g_{M}/\omega_{M} = 1.25 $ and $\Gamma/\omega_{M}=0.1$. Notice the interchange of peak heights with this change of initial phonon number.}
\end{figure}

Here we would also like to note that by setting $g_{M}$ and $\omega_{M}$ both equal to zero (i.e. neglecting the mechanical oscillations completely) we recover the usual Lorentzian spectrum emitted by a Fabry Per\'ot cavity (as one can expect and anticipate from the behaviour of top most red curve in FIG.~[2(b)]). We found that the general features of the spectrum (which we calculated using QJT) are consistent with the already present literature about the single photon optomechanical spectrum \cite{nunnenkamp2011single,rabl2011photon,ren2013single}. The spectra in these references were calculated mainly using Quantum Langevin Equations and input-output theory.

\subsubsection{Time-dependent spectrum} After finding the regime of parameters where one can detect the fully resolved single-photon optomechanical spectrum, next in FIG.~[5] we plot the time-dependent spectrum emitted by such a system.  
\begin{figure}
\includegraphics[width=7.5cm,height=5cm]{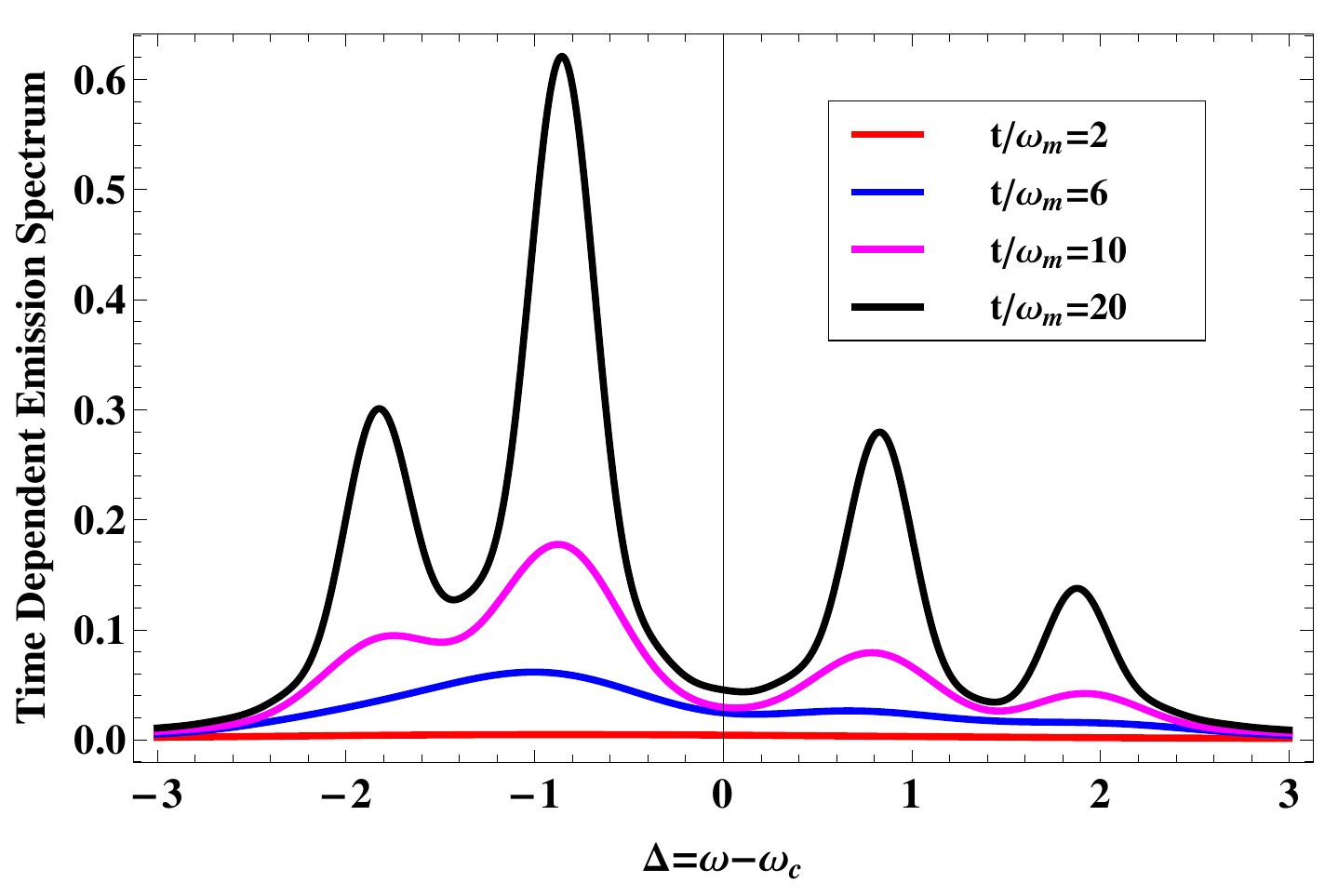}
\captionsetup{
  format=plain,
  margin=1em,
  justification=raggedright,
  singlelinecheck=false
}
\caption{Time-dependent spectrum emitted by OMC in the strong coupling regime with parameters $\kappa/\omega_{M}=0.25$, $g_{M}/\omega_{M} = 1.25$ and $\Gamma/\omega_{M}=0.1$}\label{Fig3}
\end{figure}
We note that with the passage of time the spectrum starts to grow, initially (until about $t\sim 6\omega^{-1}_{M}$) in the form of a broad curve shifted towards the $-\Delta$ axis. This situation corresponds to the times when the mechanical oscillator (which has zero phonons to begin with) has just started to vibrate and the photon leaked out before it could interact with the moving mirror. But at later times  we notice that side-bands start to emerge in the spectrum, indicating that now the photon has interacted with the mirror and has produced a phonon. Finally, after a long enough time ($t\sim 20\omega^{-1}_{M}$), the peaks become sharper and more pronounced, which is showing us directly how the mechanical oscillations have had an effect on the photon spectrum. 
\subsection{Non-zero phonon leakage and the presence of mechanical thermal bath}
In this subsection we are going to include the losses from the mechanical oscillator by assuming that the mechanical oscillator is interacting with a finite temperature Markovian mechanical heat bath with average number of thermal phonons $\overline{M}$. This coupling opens the possibility of phonons escaping from the OMC with a rate $\gamma_{M}$. Assuming that initially the heat bath and the mechanical oscillator are in thermal equilibrium, we can specify the initial thermal state of the mechanical motion by giving the probability of finding $m$ phonons as
\begin{equation}\label{IBD}
p_{m}(t_{0})=\frac{\overline{M}^{m}}{(1+\overline{M})^{m+1}}.
\end{equation} 
\begin{figure}
    \includegraphics[width=3in,height=2in]{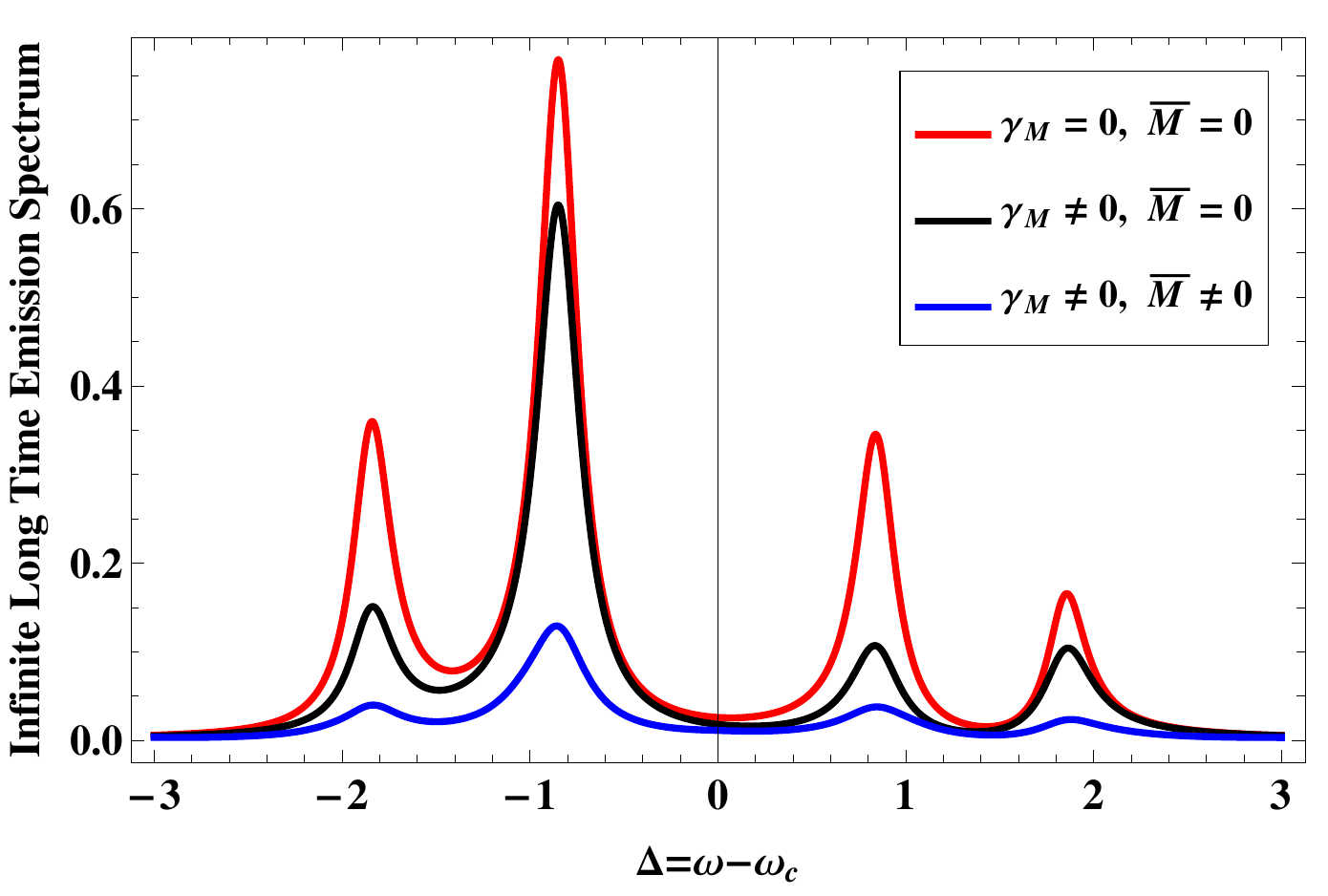}
    \captionsetup{
  format=plain,
  margin=1em,
  justification=raggedright,
  singlelinecheck=false
}
 \caption{Infinitely long time spectrum emitted by OMC including mechanical losses $\kappa/\omega_{M}=0.25$, $\gamma_{M}/\omega_{M}=0.1$, $\overline{M}=0.8$, $g_{M}/\omega_{M} = 1.25 $ and $\Gamma/\omega_{M}=0.1$.}
\end{figure}
The  spectrum in this case will be different from our previously calculated spectrum $P({\Delta,\infty})$ (Eq.[\ref{ILTS}]), and it will  be a weighted average 
over spectra calculated from different initial
numbers of phonons.
The new parameters $\gamma_{M}$ and $\overline{M}$ will enter into the calculations when identifying the non-Hermitian Hamiltonian $\hat{H}_{dNH}$, which contains some additional terms now,
\begin{equation}\label{NHHd}
\hat{H}_{dNH}=\hat{H}_{NH}-i\hbar\frac{\gamma_{M}}{2}(\overline{M}+1)\hat{b}^{\dagger}\hat{b}-i\hbar\frac{\gamma_{M}}{2}\overline{M}\hat{b}\hat{b}^{\dagger},
\end{equation}  
where $\hat{H}_{NH}$ is the same as before, as displayed in Eq.[\ref{NHH}]. Following then the same line of calculations developed in last subsection, we finally arrive at the following time-independent spectrum:
\begin{widetext}
\begin{equation}\label{ILTSD1}
\begin{split}
&P^{(1)}_{d}(\Delta,\infty)=\Bigg(\frac{1}{1+\overline{M}}\Bigg)\Bigg[\Bigg(\frac{\kappa\Gamma^{2}}{\Gamma+\overline{M}\gamma_{M}} \Bigg)\Bigg\lbrace \Bigg|\frac{i(\omega_{M}-\Delta)+\kappa/2+(\overline{M}+1)\frac{\gamma_{M}}{2}}{\lbrace i(\omega_{M}-\Delta)+\kappa/2+(\overline{M}+1)\frac{\gamma_{M}}{2}\rbrace \lbrace \kappa/2-i\Delta +\overline{M}\frac{\gamma_{M}}{2}\rbrace+g^{2}_{M}}\Bigg|^{2}\Bigg\rbrace  \\
&\hspace{17mm}+\Bigg(\frac{\kappa\Gamma^{2}}{\Gamma+(\overline{M}+1)\gamma_{M}} \Bigg)\Bigg\lbrace \Bigg|\frac{ig_{M}}{\lbrace \kappa/2-i(\omega_{M}+\Delta)+(\overline{M}+1)\frac{\gamma_{M}}{2}\rbrace \lbrace \kappa/2-i\Delta +\overline{M}\frac{\gamma_{M}}{2}\rbrace+g^{2}_{M}}\Bigg|^{2}  \Bigg\rbrace      \Bigg].
\end{split}
\end{equation}
\end{widetext}
This spectrum is shown in FIG.~[6]. For the sake of comparison we have plotted three different situations: (i) when there is no mechanical decay (red curve), (ii) when the thermal bath is at zero temperature but there is decay (black curve), and (iii) when both phonon decays and finite temperature effects are considered.

 We notice in case (ii) that there are still four peaks but the two symmetric peak heights are considerably reduced compared to other two asymmetric peaks. This fact can be explained by looking at the dressed state picture (FIG.~3). With the possibility of phonon decay, the transition between the states of one phonon to states with zero phonons will be possible. Hence the transitions (I) and (III) should be more probable now and hence peaks corresponding to these transitions (both are asymmetric) become higher than the other two symmetric peaks, which correspond to the situations in which the final state still contains a phonon. In case (iii)  all four peaks are still present and centered at the same positions, but they are now all considerably reduced in height compared to case (i). 

\section{Two-phonon mechanical oscillations}
 The more strongly the photon interacts with the movable mirror the more phonons it can generate. The restriction used in the previous Section to just a single phonon won't remain a valid assumption. With this motivation in mind, we now allow the possibility of two phonons to get an idea of what aspects of the spectrum will change with the presence of additional phonons. In this Section we present results for the case when mechanical losses are included. In the QJT approach we still have the same system (as described by the non-Hermitian Hamiltonian, Eq.[\ref{NHHd}]), but the mechanical motion annihilation operator $\hat{b}$ should now be expressed  as 
 \[\hat{b}=\ket{0}_{b}\bra{1}+\sqrt{2}\ket{1}_{b}\bra{2}\] with $\ket{m}_{b}$ is the phonon number state with $0\leq m\leq2$. 
 The restriction to at most two phonons  implies that $\hat{b}^{\dagger}\ket{m}=0$, $\forall$ $m\geq 2$. The No-Jump state  must now account for additional possibilities of finding excitations in the system, and we write
 \begin{equation}
\begin{aligned}
&\ket{\tilde{\psi}(t)} = d_{1}(t)\ket{10}+d_{2}(t)\ket{11}+d_{3}(t)\ket{12}.
\end{aligned}
\end{equation}
Here we are using a different symbol $d$ for the amplitudes just to make the distinction with the previous Section more clear. We do use the same notational convention as before so that the second slot is reserved for displaying the number of phonons, and the first slot gives the number of photons. Assuming again that initially there was no phonon in the system, the infinitely long time spectrum now has the form
\begin{widetext}
\begin{equation}\label{ILTSD2}
\begin{split}
&P^{(2)}_{d}(\Delta,\infty)=\Bigg(\frac{1}{1+\overline{M}}\Bigg)\Bigg[\Bigg( \frac{\kappa\Gamma^{2}}{\Gamma+\overline{M}\gamma_{M}}\Bigg)\Bigg|D_{1}(s=-i\Delta)\Bigg|^{2}+\Bigg(\frac{\kappa\Gamma^{2}}{\Gamma+(3\overline{M}+1)\gamma_{M}}\Bigg)\Bigg|D_{2}(s=-i\Delta-i\omega_{M})\Bigg|^{2}  \\
&\hspace{20mm}+\Bigg(\frac{\kappa\Gamma^{2}}{\Gamma+(\overline{M}+1)\gamma_{M}}\Bigg)\Bigg|D_{2}(s=-i\Delta-2i\omega_{M})\Bigg|^{2}  \Bigg].
\end{split}
\end{equation}
\end{widetext}
The expressions of the amplitudes in Laplace space are rather involved and will not be shown here. Spectrum plots (both with mechanical losses and without losses) are shown in FIG.~[7]. We note that with the inclusion of one more phonon in the system, multiple additional side bands appear. Corresponding to the Laplace amplitude $D_{1}(s=-i\Delta)$ there are three peaks now and their location can be found by setting the real part of its pole equal to zero. The remaining peaks are then located at integer multiples of $\omega_{m}$ away from these three peaks where the integer here can either be 1 or 2. 

\begin{figure}[h]
    \includegraphics[width=3in,height=2in]{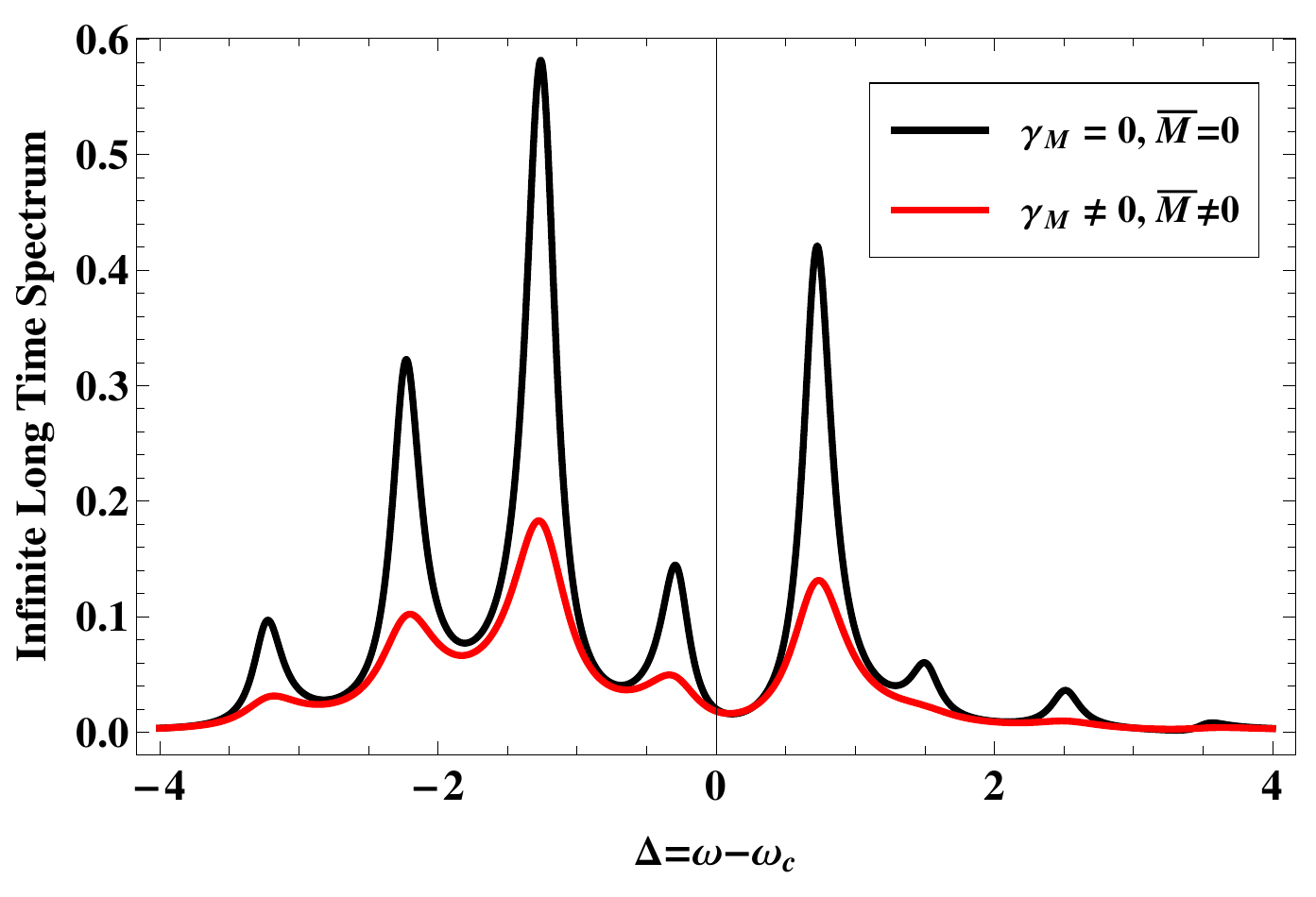}
    \captionsetup{
  format=plain,
  margin=1em,
  justification=raggedright,
  singlelinecheck=false
}
 \caption{Two-phonon infinitely long-time spectrum emitted by the OMC with mechanical losses included. Notice the appearance of many side bands when going to two phonons. Parameters are the same as in FIG.~5}
\end{figure}

In the lossless case, the analytic expression of the spectrum Eq.[\ref{ILTSD2}] predicts nine resonances, while we find eight peaks in the plot. A close inspection of the peak locations calculated from the spectrum indicates that the location of the highest peak (peak located at $\Delta = -1.25 \omega^{-1}_{M}$) occurs twice (both from $D_{1}(s=-i\Delta)$, $D_{3}(s=-i\Delta-2i\omega_{M})$). Hence the total number of peaks is one less than the naively expected number of resonances. This also explains  why this peak is the highest among all resonances.

When losses are included (red curve in FIG.~[7]) transitions among different phonon number states become more probable than before. This causes a redistribution in the peak heights, and the peaks  become wider as well. This broadening feature at finite bath temperatures also causes a dissolution of the smaller peaks.
 
Finally, we note that by following the same procedure of calculations our analysis of the  two-phonon scenario can be straightforwardly extended numerically to multi-phonon situations.
\section{Conclusions}
We demonstrated how a single photon interacts with a moving mirror, and how the single-photon time-dependent spectrum reveals the properties of that interaction. Resonances in the spectrum show how many phonons were created
and what the strength of the photon-mirror interaction is. A simple dressed-state picture 
suffices to explain the positions and relative strengths of those resonances. The time-dependent spectrum shows how the resonances are built up in time by the photon interacting with the moving mirror and generating phonons.

\bibliography{article3}
\end{document}